\documentstyle[aps,multicol,epsfig,prl]{revtex}

\begin{document}
 \input epsf
\draft \preprint{HEP/123-qed}
 \title{Fracture and second-order phase transitions.}
 \author{Y. Moreno$^1$\cite{byline}, J. B. G\'{o}mez$^2$, A.
F. Pacheco$^1$}
 \address{ $^1$ Departamento de F\'{\i}sica Te\'{o}rica, Universidad de
Zaragoza, 50009 Zaragoza, Spain.\\ $^2$ Departamento de Ciencias de la
Tierra,
 Universidad de Zaragoza, 50009 Zaragoza, Spain.}
 \date{\today}
 \maketitle
 \widetext
\begin{abstract} Using the global fiber bundle model as a tractable scheme of
progressive fracture in heterogeneous materials, we define the
branching ratio in avalanches as a suitable order parameter to
clarify the order of the phase transition occurring at the
collapse of the system. The model is analyzed using a
probabilistic approach suited to smooth fluctuations. The
branching ratio shows a behavior analogous to the magnetization in
known magnetic systems with 2nd-order phase transitions. We obtain
a universal critical exponent $\beta\approx 0.5$ independent of
the probability distribution used to assign the strengths of
individual fibers.
 \end{abstract}
\begin{multicols}{2}
\narrowtext

The interest in the fracture processes of heterogeneous media has
increased in the last years \cite{hr90,zrsv97,cb97,asl97}. In the lab,
a disordered material subjected to an increasing external load can be
studied by measuring the acoustic emissions before the global rupture.
It has been shown \cite{ggbc97,ppvac94} that this intense precursory
activity in the form of bursts of different microscopic sizes follows a
well-defined power law. In despite of the many efforts and successes
that have been recently achieved, the question of whether rupture
exhibits the properties of a first-order or a second-order phase
transition remains under discussion as well as what is the order
parameter that indicates the type of transition.

From the theory side, the understanding of fracture in heterogeneous
materials has progressed due to the use of lattice models and large
scale simulations \cite{cb97}. In this field, it is important to use
models able to describe the complexity of the rupture process,
nevertheless, they should be simple enough as to permit analytical
insights. To this class of models belong the well-known Fiber Bundle
Models (FBM) widely used since their introduction more than forty years
ago \cite{c58,d45}. In static FBM, a set of fibers (elements) is
located on a supporting lattice and one assigns to its elements a
random strength threshold sampled from a probability distribution. The
set is loaded and fibers break when their loads exceed their threshold
values. In the Equal Load Sharing (or global) FBM, which is the
simplest scheme one can adopt to make the problem analytically
tractable, one assumes that the load carried by failed elements is
equally distributed among the surviving elements of the system.

In the present Letter, we explore the criticality of fracture in
the global FBM using a novel probabilistic approach devised to
smooth fluctuations. The scaling relations obtained and the
behavior of the order parameter point out that the fracture of a
fiber set with long range interaction undergoes a continuous phase
transition.

Let us first recall the basic ingredients of the global FBM and
how one proceeds in numerical simulations. The system under
consideration is a set of $N_0$ elements located in a supporting
lattice each one having at the initial state a zero load and a
fixed strength threshold value sampled randomly from a probability
distribution $P(\sigma)$. The system is then subjected to an
external force $F$ such that each element increases its load in
the same amount, $\sigma$. This individual stress (or load),
$\sigma$, will act as the control parameter. The process of
driving is done quasi-statically; i.e., the external force is
increased at a sufficiently slow rate as to produce a single
breaking event when the stress on the weakest element equals its
threshold value. Then, the increase of $F$ stops and the load of
the broken element is equally transferred. This implies that the
load on any element is given by $\sigma=F/n_s(F)$, where $n_s$ is
the number of surviving elements for a given $F$. The rupture of
an element may induce secondary failures which in turn may trigger
more failures and so on. This process of induced failure at
constant external load, termed an avalanche, stops when all
surviving elements carry a load lower than their thresholds. The
system is then loaded again and the process is repeated until the
final catastrophic avalanche provokes the total rupture of the
material, which occurs at a critical load $\sigma_c$ that depends
on the probability distribution from where the individual
strengths were drawn, as well as on the system size. The FBM have
been recently used in self-organized criticality (SOC), a
theoretical framework widely used for the study of avalanche
phenomena in disordered systems. It has been shown using these
models that systems with plastic behavior can reach a SOC state
just before the global rupture \cite{kzh99}. A second case of
self-organization with power law distributions in several
quantities corresponds to the situation in which the fracture
process coexists with a healing process \cite{mgp99}.

In numerical simulations, the cycle of complete breakdown of the model
is performed many times in order to average out the effect of
fluctuations and obtain mean values. As we are interested in studying
the behavior of the system as the critical point, or point of final
collapse, is approached, it is of most importance to find a simple
method able to capture the evolution of the system avoiding as much as
possible the fluctuations appearing in numerical simulations. To
introduce our probabilistic strategy, let there be a large set of $N_0$
elements. Suppose that each element carries a given load $\sigma$,
which is zero at the initial state. The strength of each element is
drawn from a probability distribution $P(\sigma)$. Different
probability distributions can be considered. In materials science the
Weibull distribution is widely used,
 \begin{equation}
P(\sigma)=1-e^{-(\frac{\sigma}{\sigma_{0}})^{\rho}}, \label{eq1}
\end{equation}
 $\rho$ being the so-called Weibull index, which
controls the degree of disorder in the system (the bigger the
Weibull index, the smaller the disorder), and $\sigma_{0}$ is a
load of reference. In the following we will assume $\sigma_{0}=1$,
and therefore the loads are dimensionless. At this point, it is
worth noting that the results and the formulas derived in the
following hold for a wide class of probability distributions. We
use here the Weibull distribution for definiteness but results
have also been obtained for other distributions. Equation\
(\ref{eq1}) represents the probability that an element fails under
the individual load $\sigma$. Now, consider the case in which an
element drawn from Eq.\ (\ref{eq1}) supports a load $\sigma_1$ but
breaks under a new load $\sigma_2$. The probability that this
happens is given by
 \begin{equation}
 p(\sigma_1,\sigma_2)=\frac{P(\sigma_2)-P(\sigma_1)}{1-P(\sigma_1)}=1-e^{-(\sigma_2^{\rho}-\sigma_1^{\rho})}.
 \label{eq2}
 \end{equation}
So, the probability $q(\sigma_1,\sigma_2)$ that an element that has
survived to the load $\sigma_1$ also survives to the load $\sigma_2$
will be given by
$q(\sigma_1,\sigma_2)=1-p(\sigma_1,\sigma_2)=e^{-(\sigma_2^{\rho}-\sigma_1^{\rho})}$.

To mimic the quasi-static increase in load on the system we impose the
condition that under an external force $F$, the next breaking event
consists of one single failure. Let suppose that after the latest
avalanche, there are $N_k$ surviving elements each one bearing a load
$\sigma_k$. The new individual load $\sigma_l$ needed to provoke the
failure of just one more element is given by the solution of
$N_k-1=N_k\cdot q(\sigma_k,\sigma_l)$. Thus,
 \begin{equation}
 \sigma_l=\left[\sigma_k^{\rho}-\ln\left(1-\frac{1}{N_k}\right)\right]^{\frac{1}{\rho}},
 \label{eq6}
 \end{equation}
where in Eq.\ (\ref{eq6}) $N_k=N_0$ and $\sigma_k=0$ at the initial
state. Elevating the external force up to the $N_k\cdot\sigma_l$ level,
statistically speaking, one element breaks. As we are dealing with an
equal load sharing set, the choice of the broken element is irrelevant
because all the elements are equivalent. Once the first element fails,
the redistribution of its stress takes place. This may induce other
failures until the end of the avalanche.

Now, how many elements will survive to the situation in which $n_1$
elements with load $\sigma_1$ fail in an avalanche step? The new load
on the intact $N_1-n_1=N_2$ elements is
$\sigma_2=\frac{N_1\cdot\sigma_1}{N_2}$. So, the number $N_3$ of
elements that survive to the new load can be expressed as
\begin{equation} N_3=N_2\cdot
q(\sigma_1,\frac{N_1}{N_2}\cdot\sigma_1)=N_2\cdot q(\sigma_1,\sigma_2).
 \label{eq7}
 \end{equation}
As a consequence of applying Eq.\ (\ref{eq7}), $N_2-N_3$ elements
break and the new total number of intact fibers will support a
bigger load $\sigma_3$. The avalanche may continue and Eq.\
(\ref{eq7}) is applied again for the set of $N_3$ surviving
elements. The iterative process will stop when no new element
fails, which occurs when the right-hand side is equal to the
left-hand side in Eq.\ (\ref{eq7}). The general form of Eq.\
(\ref{eq7}) is
 \begin{equation}
 N_{j+1}=N_j\cdot q(\sigma_{j-1},\sigma_{j}),
 \label{eq8}
 \end{equation}
 with the conservation condition for the total load in the system during an
 avalanche
 \begin{equation}
 N_j\cdot\sigma_j=N_{j-1}\cdot\sigma_{j-1}
 \label{eq9}
 \end{equation}
and the condition
\begin{equation}
 N_j=N_{j+1}
 \label{eq10}
 \end{equation}
which determines the end of the avalanche. The dynamics of the system
is determined by Eq.\ (\ref{eq6}), (\ref{eq8}), (\ref{eq9}). The size
of an avalanche is given by the number of elements that break between
two successive steps of external loading. The critical load, defined as
the load needed to provoke the total collapse of the system, is equal
to the load on the intact elements just before the final catastrophic
avalanche. Note that in this probabilistic approach, in contrast to
Monte Carlo simulations, we only need to store the information
concerning the loads on the intact elements, that is, the details of
the threshold list is omitted.

In the probabilistic strategy, we can proceed in two different
ways in order to determine when an avalanche ends, to which we
will refer as the continuous and the discrete cases. For the
continuous case, the number $N_{j+1}$ of surviving elements is
considered a real number. Strictly speaking, this means that the
condition\ (\ref{eq10}) is never fulfilled. So, the condition\
(\ref{eq10}) is replaced in numerical calculations by using a
factor $\nu\ll1$ that determines the end of an avalanche, i.e., if
$N_{j}-N_{j+1}\leq\nu$ the avalanche stops; otherwise it
continues. In the discrete case, $N_{j+1}$ is considered to be a
whole number, so that after each iteration of Eq.\ (\ref{eq8}),
$N_{j+1}$ has to be rounded up. This is done comparing the
remainder of $N_{j+1}$, $\lambda$, with a random number $\alpha$
uniformly distributed in the interval $[0,1[$. Thus, if
$\alpha\geq\lambda$, $N_{j+1}$ is equal to its whole part,
otherwise, $N_{j+1}$ is equal to its whole part plus one. Next, we
check whether the condition\ (\ref{eq10}) is satisfied for the
rounded value of $N_{j+1}$ or if a new iteration of Eq.\
(\ref{eq8}) has to be performed. The continuous approach has the
advantage that the fluctuations are ruled out, whereas for the
discrete case the results are similar to those obtained by Monte
Carlo simulations where it is necessary to average over many
realizations in order to get accurate mean values. Remember that
in this model the central limit theorem applies \cite{s89}.

\begin{figure}
 \begin{center}
 \epsfig{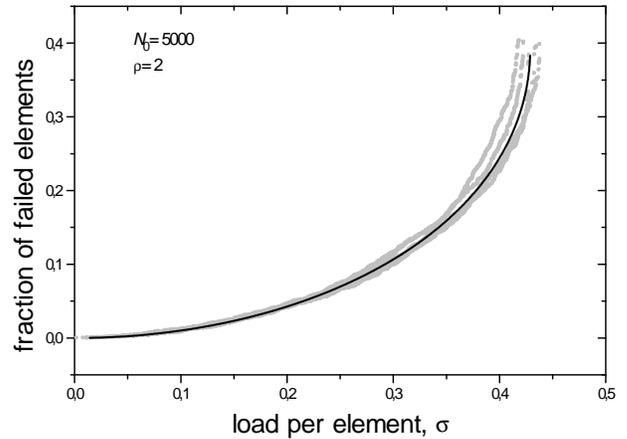}
 \end{center}
 \caption{Fraction of broken elements for the equal load sharing model. The line corresponds
 to the results obtained with the continuous approach and gray dots correspond to four Monte Carlo realizations.}
 \label{fig1}
\end{figure}

In Fig.\ \ref{fig1} we have depicted the fraction of broken
elements versus $\sigma$, for the continuous case and for four
individual Monte Carlo simulations with a Weibull index $\rho=2$
and $N_0=5000$. No averaging has been done because our aim is just
to illustrate the scatter of the results. As can be seen, the
continuous probabilistic model gives a smooth curve, and provides
an accurate value for the critical load $\sigma_c$, which
analytically is given by $\sigma_c=(\rho e)^{-1/\rho}$ \cite{d45}
in the limit of infinite $N_0$.

\begin{figure}
 \begin{center}
 \epsfig{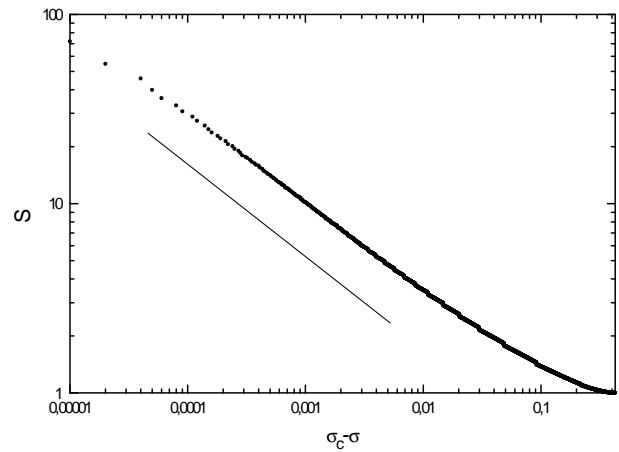}
 \end{center}
 \caption{Scaling of the mean avalanche size, $s$, as the critical point $\sigma_c$
is approached. The results correspond to the continuous probabilistic approach for a
system of $N_0=50000$ elements and $\rho=2$. The straight line with a slope
$-\frac{1}{2}$ has been drawn for comparison.}
 \label{fig3}
\end{figure}

Now, we proceed to explore the behavior of some quantities as the
critical point is reached. The results shown below have been obtained
using the continuous approach ($\rho=2$). In Fig.\ \ref{fig3} we show
the interesting scaling relation for the average avalanche size. It
turns out that the avalanche size near to the critical point diverges
with an exponent $\gamma=\frac{1}{2}$ as
$s\sim(\sigma_c-\sigma)^{-\gamma}$. A similar behavior, through a
mapping of a fuse network model to the global fiber bundle model used
here, has been recently reported \cite{zrsv97}. We have also obtained
the same scaling function for the derivative of the number of broken
fibers with respect to the load on the system. The rate $dN/d\sigma$
diverges as $(\sigma_c-\sigma)^{-\frac{1}{2}}$, thus qualifying a
critical mean field behavior as was already shown in Ref. \cite{asl97}.
In Ref.\ \cite{kzh99}, a similar scaling behavior is addressed for the
derivative of the strain carried by the fibers with respect to the
driving force.

\begin{figure}
 \begin{center}
 \epsfig{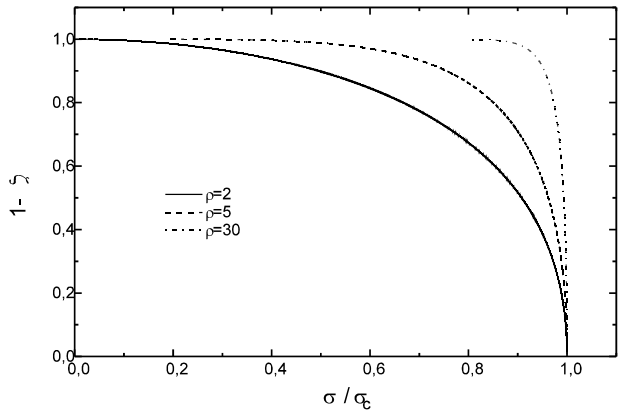}
 \end{center}
\caption{Evolution of the branching ratio as the critical point is
approached in the continuous probabilistic method ($N_0=50000$). Note
that at the critical point the branching ratio reaches the unity. The
critical exponent is $\beta=\frac{1}{2}$ which coincides, as it should,
with the value of the exponent $\gamma$.}
 \label{fig5}
\end{figure}

Another way to shed light on the critical behavior of this type of
system is to define a branching ratio $\zeta$ for each avalanche. This
magnitude represents the probability to trigger future breaking events
given an initial individual failure \cite{jensen,ccp99}, and is related
to the number of broken fibers by
 \begin{equation}
 \zeta=\frac{<z>-1}{<z>}.
 \label{eq11}
 \end{equation}
The above relation can be obtained by thinking of the evolution of
fracture as a kind of branching process \cite{h63}. In this
process, each node gives rise to a number $n$ of new branches in
the next time step. The average number $<n>$ of new branches is
called the branching ratio. Let us denote by $n_t$ the number of
branches at a given step $t$ of the branching process, and by
$t_{max}$ the total number of time steps before it dies. Then,
$\zeta=1-\frac{n_0}{\sum_{t=0}^{t_{max}}n_t}$. As, $n_0=1$,
$\zeta=1-\frac{1}{n_{tot}}$ where $n_{tot}$ is the total number of
nodes developed in the branching process. For a fracture process,
$n_{tot}$ is equal to the average number of failure events. So,
Eq.\ (\ref{eq11}) defines the branching ratio. We represent by
$<z>$ the average number of elements that fail in one avalanche,
which is a function of $\sigma$ and coincides with $s$. This
analogy between fracture and branching processes has been
previously used to study the criticality in the process of
fragmentation of Hg drops \cite{smla96}. The branching ratio will
then act as the order parameter. It takes the value 1 when the
system is critical thereby representing a measure of the distance
of the system from the critical state \cite{ccp99}. We have
numerically computed $\zeta$ by means of the continuous method.
The results obtained for a system of $N_0=50000$ elements and
several values of $\rho$ have been plotted in Fig.\ \ref{fig5}. It
can be seen in this figure that $1-\zeta$ approaches zero, in all
cases, in a continuous fashion as the critical load is reached.
Near the critical point, the relation $1-\zeta\sim
(\sigma_c-\sigma)^{\beta}$, with $\beta=\frac{1}{2}$ applies. This
is the exponent of the order parameter that we should expect from
a mean-field approach. Note the similarity of the figure with
those obtained for the magnetization in known magnetic systems
with second-order phase transitions. In the figure, the values of
$\zeta$ are collected for all the avalanches except for the last,
that which provokes the collapse of the system. The result that at
$\sigma_c$ $\zeta\rightarrow 1$, is consistent with the previous
result that the avalanche size diverges at the critical point. On
the other hand, the branching ratio does not depend on the size of
the system for large systems, in contrast with previous results in
other fracturing systems \cite{ccp99}.

The suggestion of Ref.\cite{zrsv97} is that fracture can be seen
as a first-order phase transition close to a spinodal-like
instability \cite{m94}. There, by simulating models of electric
breakdown and fracture, the authors present numerical and
theoretical evidence of several scaling relations and of a
discrete jump in some macroscopic properties. Here, we have
obtained the same scaling relation for the rate of fiber failures
as the critical point is reached, and for the avalanche sizes,
which also diverge at that point. Our numerical results also fit
the mean field result $\gamma=\frac{1}{2}$. It is true that the
fraction of unbroken fibers just before the global rupture has a
discontinuity; but from our point of view that is not enough as to
set the conclusion that fracture can be describe as a first-order
phase transition, since the concepts related to spinodal
nucleation are not sufficiently well established in driven
disordered systems.

Our alternative point of view has been to consider the {\sl
branching ratio} as an appropriate order parameter. According to
the results obtained, the branching ratio goes {\sl continuously}
from zero to one. Note, additionally, that what changes
discontinuously at $\sigma_c$ is the rate of change of $\zeta$
rather than $\zeta$ itself, which is in the essence of a
continuous phase change. Therefore, the behavior of the branching
ratio implies that the system undergoes a second-order phase
transition as claimed in other analysis of fracture models
\cite{sa98}. Our results suggest that fracture in heterogeneous
systems with long range interaction can be described as a phase
transition of the second-order type. Besides, it is important to
bear in mind that in fiber models with local interactions, the
order parameter $\zeta$, has a discontinuous jump typical of first
order phase transitions. Finally, it is worth recalling that
fracture of real materials is a process based on elasticity, and
elasticity is a long-range phenomenon. In this respect, the global
load-sharing model we have explored here could be a better analogy
to real fracture than the local one.

Y.M thanks A. Vespignani and H. J. Jensen for useful and
stimulating discussions held at the Abdus Salam International
Centre for Theoretical Physics where this work began. Y. M. is
supported by the AECI. This work was supported in part by the
Spanish DGICYT under Project PB98-1594.

\end{multicols}

\begin{references}
 \bibitem[*]{byline}  Now at The Abdus Salam ICTP, Trieste 34100,
 Italy.
 \bibitem{hr90} {\sl Statistical Models for the Fracture of Disordered
 Media\/}. Editors, H.J. Herrman and S. Roux, North Holland (1990);
 M. Sahimi, {\em Phys. Rep.} {\bf 306} (1998) 213, and
 references therein.
 \bibitem{zrsv97} S. Zapperi, P. Ray, H. E. Stanley and A. Vespignani,
 {\em Phys. Rev. Lett.} {\bf 78}(1997) 1408; {\em Phys. Rev. E.} {\bf
 59}, 5049 (1999).
 \bibitem{cb97} {\sl Statistical Physics of Fracture and Breakdown in
 Disordered Systems}. B. K. Chakrabarti and L. G. Benguigui, Clarendon
 Press, Oxford (1997), and references therein.
 \bibitem{asl97} J. V. Andersen, D. Sornette, and K.-T. Leung, {\em
 Phys. Rev. Lett.} {\bf 78}, 2140 (1997).
 \bibitem{ggbc97} A. Garciamartin, A. Guarino, L. Bellon and S. Ciliberto, {\em Phys. Rev. Lett.}
 {\bf 79}, 3202 (1997).
 \bibitem{ppvac94} A. Petri, G. Paparo, a. Vespignani, A. Alippi and M. Constantini,
 {\em Phys. Rev. Lett.} {\bf 73}, 3423 (1994).
 \bibitem{c58} B.D. Coleman, {\em J. Appl. Phys.} {\bf 29}, 968 (1958)
 \bibitem{d45} H.E. Daniels, {\em Proc. Roy. Soc.} {\bf A183}, 404 (1945).
 \bibitem{kzh99} F. Kun, S. Zapperi, and H. J. Herrmann, preprint
 cond-mat/9908226.
 \bibitem{mgp99} Y. Moreno, J. B. G\'{o}mez, and A. F. Pacheco, {\em
 Physica A} {\bf 274}, 400 (1999).
 \bibitem{s89} D. Sornette, {\em J. Phys. A} {\bf 22}, L243 (1989); J.
 Galambos, {\em The Asymptotic Theory of Extreme Order Statistics} (New
 York, Wiley, 1978).
 \bibitem{jensen} H. J. Jensen, {\em Self-Organized Criticality} (Cambridge University Press, 1998), and references therein.
 \bibitem{ccp99} G. Caldarelli, C. Castellanos, A. Petri, Physica A
 {\bf 270}, 15 (1999).
 \bibitem{h63} T. E. Harris, {\em The Theory of Branching Processes} (Berlin: Springer-Verlag, 1963).
 \bibitem{smla96} O. Sotolongo-Costa, Y. Moreno, J. J. Llovera, J. C.
 Antoranz, {\em Phys. Rev. Lett.} {\bf 76}, 42 (1996).
  \bibitem{m94} L. Monette, Int. J. Mod. Phys. B {\bf 8}, 1417 (1994).
 \bibitem{sa98} D. Sornette, J. V. Andersen, Eur. Phys. J. B, {\bf 1},
 353 (1998).
\end{references}
\end{document}